\documentclass[aps,prb,reprint,superscriptaddress]{revtex4-2}

\usepackage{graphicx}
\usepackage{amsmath}
\usepackage{amssymb}
\usepackage{float}
\usepackage[colorlinks,urlcolor=blue]{hyperref}
\usepackage[capitalise]{cleveref}   
\usepackage{nicefrac}
\usepackage{xcolor}
\usepackage{bigints}
\usepackage[normalem]{ulem}
\usepackage{multirow}
\usepackage{enumitem}

\begin{document}

\title{Shot noise signatures of candidate states for the fractional quantum Hall $\nu = 12/5$ state}

\author{Goutham Vinjamuri}
\email{vinjamuri21@iiserb.ac.in}
\affiliation{Department of Physics, Indian Institute of Science Education and Research Bhopal, Bhopal Bypass Road, Bhauri, Bhopal 462 066, Madhya Pradesh, India}
\author{Ankur Das}
\email{ankur@labs.iisertirupati.ac.in}
\affiliation{Department of Physics, Indian Institute of Science Education and Research (IISER) Tirupati, Tirupati 517619, India}

\begin{abstract}
Fractional quantum Hall (FQH) states are highly sought after because of their ability to host non-abelian anyons, whose braiding statistics make them excellent candidates for qubits in topological quantum computing. Multiple theoretical studies on the $\nu=\frac{12}{5}$ FQH state predict various quasi-particle states hosted by the $\frac{12}{5}$ plateau, which include $\mathbb Z_3$ parafermions and Majorana modes. In this work, we provide a systematic protocol to distinguish among four possible candidate wavefunctions of the $\frac{12}{5}$ plateau using zero-frequency shot noise experiments on a filter-geometry.
Qualitative comparisons of Fano-Factors provide a robust way to predict the candidate state across both the full and partial thermal equilibration regimes without prior knowledge of the experimental information, like thermal equilibration length, to allow for more realistic experiments.
\end{abstract}	

\maketitle
\section{Introduction} 

In the modern history of condensed matter physics, the quantum Hall effect and fractional quantum Hall effect have a remarkable contribution \cite{VonKlitzing1986,Tsui_fqh_1982,Halperin_fqh_1983,Laughlin_fqh_1983}. This was our first known example of a topological insulator \cite{PhysRevLett.49.405}. The fractional quantum Hall effect produces one of the most natural playgrounds of strong interaction in a topological setting. Since its discovery, one new aspect of the two-dimensional system was proposed, namely anyons which was unseen before \cite{Moore-Read1}. These anyons can be of two types --- 1) abelian (seen in most common fractions like $1/3, 2/3$, etc.), 2) non-abelian anyons (proposed to be seen in fractions like $5/2,12/5$, etc.) \cite{Wen_Non-Abelian_1991,Stern_Probe_Non_Abelian_2006}. Among the two most studied fillings corresponding to non-abelian excitations, $5/2$ has been proposed to have Majorana modes \cite{Kun2022}, and $12/5$ has been proposed to have parafermions \cite{PhysRevB.77.205310,PhysRevB.91.235112}. In the recent interest of quantum computing, these provide a playground to create quantum qubits and gates. It has been proposed that $5/2$ can help to create a NOT gate, and $12/5$ can be used to create any universal gate \cite{PhysRevB.73.245307}. However, this first requires the identification of the bulk state. There has been quite some work for the $5/2$ case. In this work, we extend the idea of Ref. \cite{SM5by2} in $5/2$ for $12/5$ and provide an exact protocol for the experimentalist to decode the bulk state purely by electrical noise measurements (The final results are summarized in \cref{Table}).

Though we discuss the principles for the conventional two-dimensional electron gas here, this idea can be extended to graphene and its sister materials. In recent times, due to its extreme controllability, we can now see more fractions \cite{science_aao2521,kumar2024quarterhalffilledquantumhall,kaur2025symmetrybrokenstateshigh,chanda2025denominatorfractionalquantumhall} that were invisible in the case of 2DEG. This protocol can provide insights into those states, too.

\begin{table}[h]
\begin{tabular}{| c | c | c || c | c | c |}
    \hline
    State & Interface & $|F^\text{Full}_{1,2,c}|$ & $F^\text{Partial}_1$   & $F^\text{Partial}_2$   & $|F^\text{Partial}_c|$\\
    \hline
    \multirow{3}{*}{HH} & $\{12/5,2\}$ & $\approx 0$ & $=0$ & $=0$ & $=0$    \\ \cline{2-6}
    &$\{12/5,7/3\}$ & $\approx 0$ &  $\approx 0$ & $\approx 0$ & $\approx 0$\\ \cline{2-6}
    &$\{3,12/5\}$ & $\approx 0.22$ &  $\approx 0.19$ & $\approx 0.19$ & $\approx 0.19$\\ 
    \hline
    \hline
    \multirow{3}{*}{BS $(N=1)$} & $\{12/5,2\}$ & $\approx 0$ & $= 0$ & $= 0$ & $=0$    \\ \cline{2-6}
    &$\{12/5,7/3\}$ & $\approx 0.072$ &  $\approx 0.22$ & $\approx 0.22$ & $\approx 0.22$\\ \cline{2-6}
    &$\{3,12/5\}$ & $\approx 0$ &  $\approx 0$ & $\approx 0$ & $\approx 0$\\ 
    \hline\hline
    \multirow{3}{*}{A-RR} & $\{12/5,2\}$ & $\approx 0.37$ & $=0$ & $\neq 0$ & $=0$    \\ \cline{2-6}
    &$\{12/5,7/3\}$ & $\approx 0.19$ &  $\approx 0.52$ & $\approx 0.52$ & $\approx 0.14$\\ \cline{2-6}
    &$\{3,12/5\}$ & $\approx 0  $ &  $\approx 0.18$ & $\approx 0.18$ & $\approx 0.18$\\ 
    \hline\hline
    \multirow{3}{*}{BS $(N=-3)$} & $\{12/5,2\}$ & $\approx 0.78$ & $=0$ & $\neq 0$ & $=0$    \\ \cline{2-6}
    &$\{12/5,7/3\}$ & $\approx 0.36$ &  $\approx 0.41$ & $\approx 0.41$ & $\approx 0.09$\\ \cline{2-6}
    &$\{3,12/5\}$ & $\approx 0$ &  $\approx 0.21$ & $\approx 0.21$ & $\approx 0.21$\\ 
    \hline
\end{tabular}
\caption{Collection of all calculated Fano-Factors for three device set-ups having $\{\nu,\nu_i\}$ pairs out of $2,3,7/3$ with $12/5$ such that $\nu>\nu_i$. $|F_{1,2,c}^\text{Full} |$ represents the fano factor in full thermal equilibration, where auto-correlations at drains one and two and cross-correlations are all equal. $F_1^\text{Partial}, F_2^\text{Partial}$ represent auto-correlation fano factors at drains one and two, respectively, in partial thermal equilibration, and $|F_c^\text{Partial}|$ represent cross-correlation in partial thermal equilibration regime. The values for fano factors at drain two for A-RR state in $\{12/5,2\}$ and for BS $(N=-3)$ in \{12/5,2\} were not calculated in this work as they are non-zero but non-universal with a strong dependence of different parameters of the experimental details (see Ref. \onlinecite{Park2020}). }
\label{Table}
\end{table}

\subsection{Details on the $\nu=12/5$ FQH state}
One of the most prominent candidates for a topological qubit is $\nu=12/5$. This is because there are predictions of one of the possible states to hold what are called the Fibonacci anyons \cite{PhysRevB.59.8084}. However, as of now, the numerical evidence is hardly prominent. Though from prior experiences with $\nu=5/2$ \cite{Banerjee2018,Dutta2022,Dutta2022_Iso,PhysRevLett.80.1505, PhysRevLett.84.4685, PhysRevLett.101.016807, PhysRevLett.105.096802, PhysRevLett.104.076803, PhysRevLett.106.116801, PhysRevX.5.021004,PhysRevLett.100.166803, PhysRevB.79.115322}, one has to take the limitations of the limited system size numerical calculations with a grain of salt. There are many candidates for the 12/5 wavefunction, but we will focus on only 4 of them. We will consider four different cases, namely Haldane Heirarchy states (HH), Bonderson-Slingerland states (BS) with $N=1$ and $N=-3$ , anti Read-Rezayi states (A-RR) \cite{PhysRevB.108.L241102}. 
\begin{figure}
    \centering
    \includegraphics[width=\linewidth]{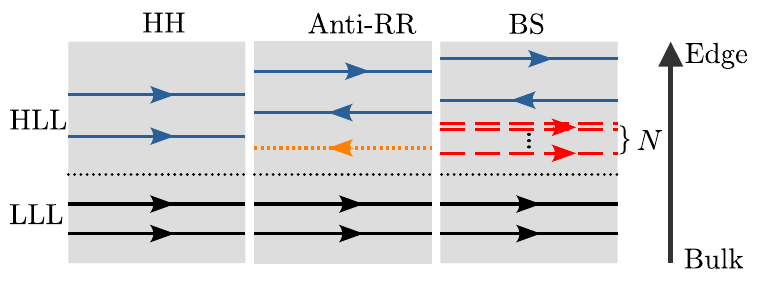}
    \caption{A representation of the edge structure for all candidate states considered in this work. The black dotted line separates the Lower Landau Levels (LLL) from the Higher Landau Levels (HLL). The thick black lines represent the Integer modes with $\nu=1$ and thick blue lines represent fractional modes with $\nu=1/5$. The dotted orange line represents $\mathbb Z_3$ parafermion modes and the dashed red lines represent Majorana modes. The BS state has two variants which we consider where the number of Majorana modes can either be $N=1$ or $N=-3$ where the negative means that they states are moving upstream. It is important to note that while the overall composition of the constituent edge modes is accurately represented by the figure, it gives no sense of which modes are facing the bulk.} 
    \label{fig:edgestructure}
\end{figure}
The $N$ in the BS states represent the number of Majorana modes in the edge structure, where the sign represents direction of the modes. $+(-)$ is downstream(upstream). There have been previous attempts to distinguish these candidate states via upstream noise. However, this completely neglects the fact that it is extremely hard and needs different geometry to check equilibration length \cite{Srivastav2021, Melcer2022, Kumar2022, Srivastav2022}. In this work we focus on shot noise and one specific geometry with an algorithmic approach where we provide a flow chart for the experiment to follow to clearly identify the details. This geometry and these ideas of measuring shot noise were shown for both abelian cases and non-abelian cases as well.

\subsection{Shot Noise}
Shot-noise measurements happen to be one of the most unique and strongest methods of measurement that gave us one of the two most important pieces of information about the system \cite{Dutta2022,
Dutta2022_Iso}. Shot-noise is one of the direct methods to measure the charge of the carrier, and it can be used to make one of the best, if not the best, electron thermometers. In the case of the conventional quantum point contact (QPC), the shot noise is generated by the discrete transition of the carrier \cite{manna2024shot}.  We quantify the noise at QPC using Fano Factor (FF), which is defined using the time averaged zero frequency noise at the drains, 
\begin{align}
F_{i,j} = \frac{S_{ij}}{2e I t(1-t)},
\end{align}
where, $   S_{ij} = \langle \delta I_i \delta I_j \rangle$ and $\delta I_i = I_i - \langle I_i \rangle$. $\delta I_i$ represents the variation in current at drain $i$ (refer to \autoref{fig:combinedfigure}), and $S_{ij}$ represents the auto and cross correlations in current at the drains for $i=j$ and $i\ne j$. When $i$ and $j$ are equal (to either 1 or 2), we calculate current-current auto-correlation at drain $i$, giving us auto-correlation fano factor. When they are not equal, we calculate current-current corss-correlation and the cross-correlation fano factor (In \cref{Table} $F_{ij}$ is shortned to $F_i$ for $i\in\{1,2,c\}$, representing current-current auto-correlation at drains 1 and 2, and current-current corss-correlation betwen drains 1 and 2). $I$ is the current entering the source and $t$ is the fraction of the current making it past the QPC to drain. We use the same formula but here $I$ is the current from the source but $t$ is the current exiting to the drain from the interface. However, the Johnson-Nyquist noise plays a very important role in our case. Specifically, one needs to measure the scaling of the noise with the source potential, giving us the Fano factor, which we will compare. The method we will use needs Johnson-Nyquist noise, which was first discussed in this context to understand shot-noise in a QPC plateau \cite{Bhattacharyya2019,Bid2010}. After the charge and thermal equilibration, if heat propagates parallel to the current (ballistic), no extra Johnson-Nyquist noise will be created. On the contrary, if the heat can propagate anti-parallel to the current (anti-ballistic), then that can create extra noise. With this information, we can calculate different Fano Factors and compare them for their qualitative behavior to narrow down the phase of the quantum Hall states.

\subsection{Device and results}

We propose the design of a heterostructure device as shown in \cref{fig:combinedfigure} \cite{Christian2020,SM5by2}, which contains two different filling fractions in an arrangement that creates interfaces (filter-geometry) where edges of different FQH states interact by a virtue of the geometry of the setup. The protocol we present to distinguish between candidate states of the 12/5 FQH state systematically involves performing the experiment with three different combinations of the pair $\{\nu,\nu_i\}$. $\nu$ is the background filling near the source/drain and $\nu_i$ is the filling in the middle region under the gate. The pairs we consider are $\{12/5,2\}$, $\{12/5,7/3\}$, $\{3,12/5\}$
to conclusively determine the candidate state in any thermal equilibration regime (partial vs full). There has not been any extensive study about understanding different internal length scales (e.g., charge equilibration length $l_\text{eq}$, thermal equilibration length $l_\text{eq}^Q$) and their order of magnitude for $12/5$ specifically. However, there have been studies that indicate that fractional states take longer to equilibrate than integers \cite{Dutta2022_Iso,Banerjee2017, Srivastav2021,Melcer2022}.

$l_\text{eq} \ll l_\text{eq}^Q$: Previous works on $\nu=2/3$ and $\nu=5/2$ suggest that these lengths are different by order of magnitude, suggesting that setups are almost always charge-equilibrated as the required geometric length for that is orders of magnitude smaller than the scale of the setup \cite{Srivastav2021,Melcer2022,Dutta2022_Iso}. Thus, we make the realistic assumption that this behavior holds even for the $\nu=12/5$ states and thus exploit it by assuming that all geometric edge lengths in our setup are charge-equilibrated by default. Thermal equilibration length is a different matter still, as the ranges of $l_\text{eq}^Q$ vary to such a non-trivial degree that shot-noise experiments can distinguish between no, partial, and complete thermal equilibration.
As such, the methodology we propose in this work (\cref{sec:method}) assumes no knowledge of thermal equilibration (we ignore no thermal equilibration in this discussion) while providing a robust and systematic elimination-based approach to determine the state.

\subsection{Plan of the paper}

In \cref{sec:Discuss} we discuss the main calculations and assumptions of the problem. We explain the location and the consequences of hotspots and noise spots as measured correlations in the current. In \cref{sec:method}, We describe the protocol for the experimentalists to distinguish different possible phases with a given flowchart. In the final \cref{sec:outlook}, we discuss the outlook and experimental feasibility.

\section{Discussions on calculations}
\label{sec:Discuss}
\subsection{Hotspots and noise-spots}

\begin{figure}
\centering
\includegraphics[width=\columnwidth]{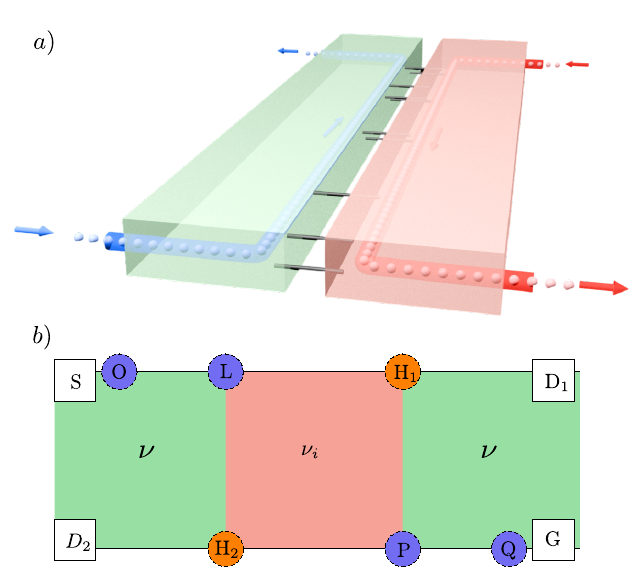}
\caption{(a) Illustration of an interface consisting of two different filling fractions such that the upstream and downstream modes contain different edge structures is modeled as shown. The chiral modes are shown in blue and red in the filling fractions $\nu$ and $\nu_i$, respectively. The grey lines between the two sides show the tunneling channels between the upstream and downstream modes to facilitate equilibration. (b) A schematic of the device used in this work. Here, contacts $S, D_1, D_2$, and $G$ are Source, 2 drains, and ground, respectively. The convention $\nu > \nu_i$ is chosen so that the current always splits by the interface.
Points labeled $O,L,P,Q$ represent noise spots, and points $H_1,H_2$ are hot spots.}
\label{fig:combinedfigure}
\end{figure}

We begin with an overview of the model used, which describes regions that have an elevated localized temperature due to power dissipation or the ``hotspots" and regions that generate a random fluctuation in charge carriers called the ``noise spots" \cite{Christian2020, Christian2019, Park2019, Hein2023}. Each hydrodynamic mode is labeled by two quantities $\nu$ and $c$, which are the filling fraction and Central Charge (CC), respectively. The filling fraction determines the quantized electrical current carried by the edge mode and central charge, the quantized thermal current \cite{Kane1997}, given by   
\begin{align}
I = \nu \frac{e^2}{h} V, \hspace{0.5cm} J = c\frac{\pi^2 k_B^2}{6h}T^2\,, \label{eq:hallcurr}
\end{align}
where $V$ is the edge voltage, $h$ is the Planck's constant, $k_B$ is the Boltzmann constant and $T$ is the temperature of the hotspot.
A candidate state can have multiple modes traveling either upstream or downstream (with respect to electrical current) while the effective filling fraction ($\nu_\text{eff}$) is positive (we chose the downstream modes to have positive FF and the upstream modes to have negative FF). We use $\nu_\text{eff}$ instead of $\nu$ in order to extend our language to different equilibration regimes. In partial equilibration, when the lower Landau levels equilibrate and nullify at the interface, the resultant conductivity is only contributed by the higher Landau levels which is given by the filling fraction $\nu_\text{eff}$. We label the upstream modes by $\nu_+$ and the downstream modes by $\nu_-$ with $c_+$ and $c_-$ as their respective central charges. In this model, three situations can arise where $c_+ > c_-$ (thermal current is in the direction of electric current) or $c_- > c_+$ (thermal current is in the opposite direction of electric current), or $c_+=c_-$ (no thermal current) \cite{Christian2020}. To use this model we must first identify the places in our device which have the same voltages, and hence no joule heating effects (which will act as our noise spots). We do this using simple current conservation along the edges of our device, (\cref{fig:combinedfigure}) which tells us the relation between voltages at certain points on the device.
\begin{subequations}
\begin{align}
\nu \frac{e^2}{h}V_O = \nu_i \frac{e^2}{h}V_L + (\nu-\nu_i) \frac{e^2}{h}V_L ,\\
\nu \frac{e^2}{h}V_Q = \nu_i \frac{e^2}{h}V_P + (\nu-\nu_i) \frac{e^2}{h}V_P.
\end{align}
\end{subequations}
This tells us that the points O, L, and, P, Q are at the same voltages.

We now set up the considerations and assumptions used in our model.
We assume $N$ equidistant `virtual reservoirs' across the length of the edge, which act as gauges for the voltages and currents at each point along the edge. Each hydrodynamic mode equilibrates with another by charges (electric and thermal) tunneling across from one mode to the other. We also assume that the reservoirs themselves do not accumulate electric or thermal currents given by, 
\begin{subequations}
\begin{align}
I_{j+1,n} = I_{j,n} + \sum_{\substack{i=1\\i\ne n}}^N I_{j,n,i}^\tau,\label{eq:eleCons}\\
J_{j+1,n} = J_{j,n} + \sum_{\substack{i=1\\i\ne n}}^N J_{j,n,i}^\tau.\label{eq:thCons}
\end{align}
\end{subequations}
Here $I(J)_{j,n}$ represents electric (thermal) current at reservoir site $j$ of mode $n$, and $I (J)_{j,n,i}^\tau$ represents tunneling electric (thermal) current from mode $n$ to mode $i$ at site $j$. The framework for computing the voltage profiles along this edge with multiple modes has been previously derived in \cite{Spanslatt2019, Park2019} and subsequently used in \cite{Christian2020,Park2020} to classify edge states using shot noise. But these works do not take the different thermal equilibration regimes possible in a device of this design, especially for a state like $\nu=12/5$ where there are no experimental estimates of thermal equilibration length. Despite that fact, one can appreciate the fact that the methodology to calculate voltage profiles and temperatures at the noise-spots does not change. Therefore we provide an overview of the formulation used to compute the necessary quantities in the subsequent sections.

The tunneling currents at a site are coupled directly to the voltage and temperature differences at the site as

\begin{subequations}
\begin{align}
I_{j,n,i}^\tau &= g\frac{e^2}{h}\Delta V_{j,n,i}\label{eq:eletunnelling},\\
J_{j,n,i}^\tau &= g\frac{e^2}{2h}\Delta V^2_{j,n,i} + \gamma g \kappa \Delta T^2_{j,n,i}, \label{eq:thtunnelling}
\end{align}
\end{subequations}
where $\Delta Q_{j,n,i}$ represents difference between quantity $Q$ in modes $n$ and $i$ at site $j$. $\gamma$ represents the factor by which ratio between electric current and thermal current and tunneling between two modes deviates from the Lornez number in Wiedemann-Franz law. In the calculations it is taken to be one \cite{roy2025halfintegerthermalconductanceabsence}. From \cref{eq:hallcurr,eq:eleCons,eq:thCons,eq:eletunnelling,eq:thtunnelling} we arrive at a set of linear differential equations for voltage and temperature at each reservoir site and mode $j, n$ which can be compactly represented matrix differential equations.
\begin{subequations}
\begin{align}
\partial_x V(x) &= \Lambda_V(g,\gamma) V(x)\label{eq:voltageDiff},\\
\partial_x T^2(x) &= \Lambda_T(g,\gamma)T^2(x) + \Lambda_J(V).\label{eq:tempDiff}
\end{align}
\end{subequations}
Here, $\Lambda_J(V)$ is a result of the Joule heating and hence is a function of the voltages.  Solving \cref{eq:voltageDiff} for voltage profiles of the modes as a function of $x$ using boundary condition $V(0) = V_0$, we
show that the potential along the interface edge is constant throughout the length, except towards the end, where a sharp voltage drop is observed \cite{Christian2020}.

The drop in power because of this drop in voltage causes the downstream end of the edge to heat up, which is referred to as a Hot spot. The hot spot can now act as a thermal source for any mode capable of carrying thermal current. \cref{fig:combinedfigure} shows two interfaces and their corresponding hot spots labelled $H_1$ and $H_2$. The device is symmetric under inversion, so we can focus on any one interface because the physics is identical at both of them. Consider the interface $LH_2$; We assume that the chirality of the edge modes in our device is clockwise. Upstream modes along the segment $LH_2$ can carry heat from $H_2$ towards $L$. As the heat travels upstream, the thermal charge carriers can either tunnel across over to the downstream mode towards $H_2$ or reach $L$, which is an entirely random process and can be affected by the length of the edge to allowing for different degrees equilibration. Therefore, if the edge $LH_2$ is such that the net thermal current is in the upstream direction (anti-ballistic), then at point $L$, one would see fluctuations in the thermal charges reaching it because of random tunneling processes along the length of the edge. One can quantify the randomness and fluctuation in thermal charges reaching $L$ by calculating Current-Current auto correlation functions and defining a fano-factor at this junction, which we will see in the next section. 

Although junctions $L$ and $P$ are noise-spots, we can only measure the noise at the
contacts of the device, which are $S$, $D_1$, and $D_2$. By exploiting the rotational
symmetry in the device, we can determine the presence and absence of noise by only looking at
the contact $S$ and the junction adjacent to it. The segments $OL$, $LH_2$ and $LH$
contribute to transporting thermal charge towards $S$. So, in the situation that the thermal
charge transport is Ballistic in all three segments, we can conclude that $S$ will experience
no noise. We construct an array ($[OL,LH_2,LH_1]$) to represent the nature of flow in these
three segments while discussing individual cases in later sections, where each element tells
the nature of thermal current flow in the corresponding segment, namely ballistic (B), anti-ballistic (AB).

Before proceeding to the calculations it is worth noting that the descripiton of the number of noise-spots and hot-spots as shown in \autoref{fig:combinedfigure} is in fact incomplete. Because $H_1$ and $H_2$ are not the only places where the voltage drops, causing joule heating. Since the drains are grounded, they experience a voltage drop as well, causing two more hotspots at the drains which may supply heat to $L$ and $P$. In our treatment of this problem we do not consider the effects of these hotspots because of a couple of considerations. As seen in later sections and in previous works \cite{Granger2009, manna2024shot}, the contacts are assumed to be in equilibrium with the surroundings (assumed to be at zero temperature) and will absorb additional heat from the new hotspots minimizing any thermal fluctuation noise that might be generated. Therefore we believe that the change in temperature at the drains due to the hotspots would be singnificanlty lower in comparission to the hotspots at $H_1$ and $H_2$ for them to have an effect on the noise at $L$ and $P$. That being said, it is not feasible to maintain the device and the contacts at absolute zero, and hence the assumption that heat generated at the drains will be absorbed by the contacts may not be realistic. Previous works \cite{Bid2010, Bhattacharyya2019, Christian2020} implement a three-contact arrangement at the drain, which prevents heat from propagating towards $L$ and $P$ via upstream modes, which is an experimentally feasible solution to avoid additional hotspots. This arrangement involves a terminal grounded contact preceeded by two floating contacts, in such a way that the contact closest to the ground acts as a heat reservoir, preventing any upstream propagation of heat.

\subsection{Auto and cross correlation fano-factors}

The goal of this section is to provide an outline of the calculations for auto and cross current-correlations, but we highly recommend going through \cite{Christian2020} for more details. Measurable quantities in an experiment for a device as shown in \cref{fig:combinedfigure} are the variances in current measured at both the drains $D_1$ and $D_2$. If current $I_1$ and $I_2$ end up at the respective drains $D_1$ and $D_2$, then the variances can be given by

\begin{subequations}
\begin{align}
    \delta I_i &= I_i - \langle I_i \rangle\\
    S_{ij} &= \langle \delta I_i \delta I_j \rangle.
\end{align}
\label{eq:variances}
\end{subequations}
% where we call $\delta^2 I_i$ (for $i=1,2$) as auto-current correlations and $\delta^2 I_c$ as cross-current correlation. 

Here for both $i$ and $j$ equal, $S_{ij}$ calculates the current-current autocorrelation at drain $i$; else we calculate the cross-correlation between drains $i$ and $j$.
We can call the quantities $\delta I_i$ as fluctuations in the respective currents, which are given by net fluctuations in the segments leading up to the drains given by $\Delta I_{LH_1} + \Delta I_{H_1 P}$ and $\Delta I_{LH_2}+\Delta I_{H_2 P}$ respectively for $i=1,2$. Note that $\delta I_i$ represents variation of $I_i$ from its mean, and $\Delta I_i$ represents change in the quantity $I_i$. Current fluctuations in the system are primarily because of voltage fluctuations and thermal fluctuations, which are as follows. 
\begin{subequations}
\begin{align}
\Delta I_{LH_1} &= \nu_i \frac{e^2}{h}\Delta V_L + \Delta I_{LH_1}^Q, \\
\Delta I_{LH_2} &= (\nu-\nu_i)\frac{e^2}{h}\Delta V_L + \Delta I_{LH_2}^Q,\\
\Delta I_{H_1 P} &= (\nu-\nu_i) \frac{e^2}{h}\Delta V_P + \Delta I_{H_1P}^Q,\\
\Delta I_{H_2P} &= \nu_i \frac{e^2}{h}\Delta V_P +\Delta I_{H_2P}^Q.
\end{align}
\end{subequations}

Using these quantities in \cref{eq:variances} along with the Johnson-Nyquist equations, which provide variance in thermal fluctuations as functions of temperature, we find the auto and cross correlation equations. 

To complete the calculation, we need the noise at the source (O) and ground (Q) (\cref{fig:combinedfigure}) and temperatures of the hot-spots which are dependent on the nature of transport in each of the three segments. Because they determine the direction of electric and thermal current leading to different conservation equations in each case. 
The noise at O and Q is given by \cite{Christian2020}
\begin{widetext}
\begin{align}
S = \frac{2e^2}{h l_\text{eq}}\frac{\nu_-(\nu_+-\nu_-)}{\nu_+}\int^L_0dx \frac{e^{-\frac{2x}{l_\text{eq}}}k_B(T_+(x) - T_-(x))}{[1-(e^{-\frac{L}{l_\text{eq}}}\frac{\nu_-}{\nu_+})]^2}\label{eq:noiseIntegral}
= \frac{e^2\nu_\text{eff}k_B T_L\nu_-}{h\nu_+}\left[\frac{\sqrt \pi \Gamma(\frac{2+\alpha}{\alpha})}{2\Gamma(\frac{4 + 3\alpha}{2\alpha})} + {}_2F_1\left(-\frac{1}{2},\frac{2}{\alpha};\frac{2+\alpha}{\alpha};\frac{c_+}{c_-} \right)\right].
\end{align}
\end{widetext}
Where $\alpha = -(n_+ - n_-)\gamma\frac{\nu_+\nu_-}{\nu_\text{eff}}$ and $\nu_\text{eff}$ is the effective filling fraction, which is important in case of partial equilibration. The length $l_\text{eq}$ is the charge equilibration length for the $\nu=12/5$ sample and $L$ is the geometric length of the edge. $T_{\pm}(x)$ is the temperature of the downstream or upstream mode as a function of $x$ along the interface of length $L$. $T_+$ and $T_-$ are calculated by solving \autoref{eq:tempDiff} by fixing the temperature $T_L$ at $x=L$ and $0$  at $x=0$ as boundary conditions. $T_L$ is the temperature of the noise-spot `L' which is at one end of the interface.

\subsection{Case-by-case discussion of each state}
After having developed the necessary tools to perform Auto and Cross-correlation calculations, we can now talk about how to calculate them. In this work, we have proposed the use of three different filling fractions in combination with $12/5$ to completely distinguish among the candidate states, which are $\{\nu,\nu_i\} =$\{12/5,2\}, $\{12/5,7/3\}$, and $\{3,12/5\}$. Unlike in the $\nu=5/2$ case \cite{MA2024324}, for $\nu=12/5$ case, experimental or numerical studies on the thermal equilibration lengths and regimes are not abundant. Therefore, to prepare a robust methodology to reliably predict the candidate state, we need to be able to identify the equilibation regime from the noise in the system. But as it turns out in the case of $\nu=12/5$, for the most part, one can bypass the need of knowing the thermal equilibration regime to distinguish between candidate states, as we will show in a later section. The following discussion also needs one to remember our assumption that all downstream modes equilibrate with other downstream modes and similarly upstream modes equilibrate with other upstream modes (in partial equilibration, upper and lower Landau levels do not equilibrate, but in full equilibration they cannot be distinguished).

\textit{Haldane Heirarchy:} In full thermal equilibration the effetive central charge of HH is ${c_\text{eff}} = 4$ ($2+2$ from lower and higher LLs). The $\nu_i = 2$ interface shows $c_{SN_1} = 4$, $c_{N_1H_2} = 4-2 = 2$ and $c_{N_1H_1} = 2$, resulting in the transport to be [B,B,B], giving us exponentially suppressed noise. $\nu_i=7/3$ contributes a net central charge of ${c_\text{eff}} = 3$ ($2+1$ from $\nu=2$ lower LL and $\nu = 1/3$ upper LL). So the segments have $c_{SN_1} = 4$, $c_{N_1H_2} = 4-(2+1) = 1 = 2$ and $c_{N_1H_1}  =3$ also showing [B,B,B] transport and exponentially suppressed noise. The $3$ filling shows a net central charge ${c_\text{eff}}=3$, giving us $c_{SN_1} = 3$, $c_{N_1H_2} = 3-4=-1$ and $c_{N_1H_1}=4$ this time resulting in [B,AB,B] transport, which means that we expect a constant non-zero fano-factor in this situation. 

In partial thermal equilibration HH state only contributes a net central charge ${c_\text{eff}}=2$ from the higher LL. At the $\nu_i=7/3$ interface, we see $c_{SN_1}= 2$, $c_{N_1H_2} = 2-1=1$ since only the $1/3$ charge mode contributes to the central charge from lower LL, and $c_{N_1H_1}=1$ giving us a net [B,B,B] transport. In the case of the filling $3$ interface $c_{SN_1}=1$ and $c_{N_1H_1}=2$ but $c_{N_1H_2}=1-2=-1$, giving us [B,AB,B] transport. 

\begin{figure*}
\includegraphics[width=\textwidth]{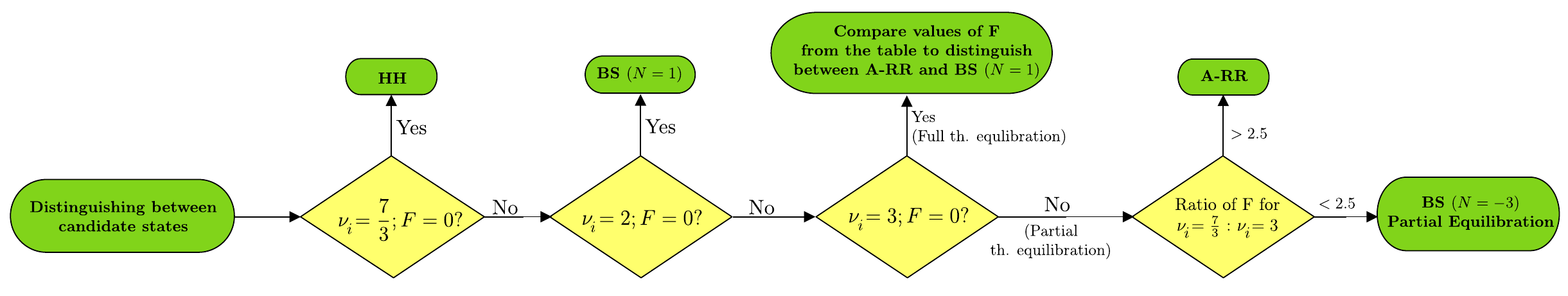} 
\caption{The flowchart shows step-by-step our protocol explained in \cref{sec:method}. Each decision point helps eliminate or confirm a state entirely based on qualitative comparisons of Fano-Factor values, bypassing the need to know the exact values of these Fanofactors.}
\label{fig:flowchart}
\end{figure*}

\textit{BS-MR ($N=1$):} Like HH, the $N=1$ state is also a particle-like state showing a net central charge ${c_\text{eff}}=5/2$ in full thermal equilibration (2 integers in the lower LL and a Majorana fermion in the higher LL). With the $\nu_i=2$ interface we can see that $c_{SN_1}=5/2$, $c_{N_1H_2}=5/2-2=1/2$, and $c_{N_1H_1}=2$. Therefore the behavior of $\nu_i=2$
is again [B,B,B], which means the noise will be exponentially suppressed. Simillary for the $\nu=3$ filling we see a completely ballistic transport as $c_{SN_1}=3$, $c_{N_1H_2}=1/2$ and $c_{N_1H_1}=5/2$. With $\nu_i=7/3$ $c_{SN_1}$ and $c_{N_1H_2}$ are ballistic with central charges $5/2$ and $3$ respectively, but $c_{N_1H_2}$ shows anti-ballistic transport with central charge $-1/2$ giving us a [B,AB,B] transport having a constant fano-factor.

Even in the partial thermal equilibration regime, the overall behavior of the BS-MR state remains the same. The $\nu_i=2$ 

For $\nu_i=7/3$ $c_{SN_1}=1/2$, $c_{N_1H_2}=-1/2$ and $c_{N_1H_1}=1$ giving us a [B,AB,B] which corresponds to a constant fano-factor. In the case of $\nu=3$ we see that $c_{SN_1}=1$, $c_{N_1H_2}=1/2$ and $c_{N_1H_1}=1$ showing completely ballistic transport ([B,B,B]) and hence exponentially suppressed noise.

\textit{Anti-RR}: The Anti Read-Rezayi state is composed of two integer modes in the lower Landau levels and 2 counter-propagating bosonic modes with a $\mathbb Z_3$ parafermion, contributing a net central charge of ${c_\text{eff}}=2-4/5=6/5$. In full thermal equilibration, with the $\nu_i=2$ interface we see $c_{SN_1}=6/5$ and $c_{N_1H_1}=2$ but $c_{N_1H_2}=2-4/5-2 = -4/5$, giving us [B,AB,B] transport which translates to a constant fano-factor. Similarly, in the $\nu_i=7/3$ case, we see $c_{SN_1}=6/5$ and $c_{N_1H_1}=2$ with $c_{N_1H_2}=2-4/5-3 = -9/5$ which is again [B,AB,B]. But in the case of $\nu=3$ we see a completely ballistic transport in all three segments ([B,B,B])as $c_{SN_1}=3$, $c_{N_1H_1}=9/5$ and $c_{N_1H_2}=2-4/5 = 6/5$.

In partial equilibration, the effective contribution of central charge is ${c_\text{eff}}=-4/5$.
With $\nu_i=7/3$ the effective central charge is only ${c_\text{eff}}=1$, so $c_{SN_1}=-4/5$, $c_{N_1H_2}=-4/5-1=-9/5$ and $c_{N_1H_1}=1$ resulting in a [AB,AB,B] transport which also means constant CCC. And for $\nu=3$ the net central charge is ${c_\text{eff}}=3$, which means $c_{SN_1}=1$ and $c_{N_1H_1}=9/5$ but $c_{N_1H_2}=-4/5$ resulting in constatnt fano-factor.

\textit{BS $(N=-3)$:} Like the BS-MR ($N=1$) state, the BS state with $N=-3$, also known as the BS-aPf state, exhibits two counterpropagating bosonic modes with the difference of $3$ Majorana modes in the upstream direction (excluding the two lower integer modes). In full thermal equilibration, the net central charge contributed by BS-aPf is ${c_\text{eff}}=2-3/2=1/2$. The $\nu_i=2$ and $\nu_i=7/3$ both show [B,AB,B] transport with central charges $[1/2,-3/2,2]$ and $[1/2,-5/2,3]$ respectively. And the $\nu=3$ shows exponentially suppressed noise with the three segments having central charges $[3,5/2,1/2]$.

In the partial equilibration regime, the effective contribution to central charge by the BS-aPf state is just ${c_\text{eff}}=-3/2$ as the counter-propagating bosonic modes equilibrate with each other, and we are left with the upstream Majorana modes. Hence, $\nu_i=7/3$ interface now shows [AB,AB,B] behaviour with central charges $[-3/2,-5/2,1]$. Finally the $\nu=3$ filling shows $[B,B,AB]$ behaviour with charges $[1,5/2,-3/2]$.

In the discussion until now, we have not discussed the partial equilibration cases for any state in the $\nu_i=2$ interface. This is because the Fano-Factor computations in this situation discussed earlier need modification as the way noise is generated is slightly different. Effective contribution of filling fraction and central charge in partial equilibration regime only come from higher Landau Levels, since higher and lower Landau Levels are not allowed to mix completely. This results in no electric or thermal conduction in the $\nu_i=2$ part of the device. Therefore no current variation is observed at $D_1$ when $\nu_i=2$ in partial equilibration. $D_2$ will only experience fluctuation in charges if the edge hosts upstream modes to carry charge carriers away from the drain. Using this analysis we observe that $F_1^\text{partial}=F_c^\text{partial}=0$ all the time. The particle-like states HH and BS $(N=1)$ host no upstream modes, therefore $F_2^\text{partial}$ is zero for them. To calculate the remaining two FFs one needs to numerically solve \cref{eq:noiseIntegral} by using the temperature profiles from \cref{eq:tempDiff}. In the case of non-abelian states hosted by $\nu=5/2$, \cite{Park2020} show the non-universal character of zero frequency noise given by $S = c_1 -c_2\sqrt{L/{l_\text{eq}}}$, where $L$ is the length of the edge and ${l_\text{eq}}$ is the charge equilibration length. \textit{Park et al.} suggest that the constants $c_1$ and $c_2$ can be estimated by comparing with the current entering the source and by finding at which values they observe constant $S$ as $L/{l_\text{eq}}$ is varied. In this work, we do not do the above analysis because the high variability of $S$ in this case likely means that it is a bad metric to be used in our protocol for accurate prediction of candidate state based on this method.
In Ref. \onlinecite{SM5by2}, the fano factor mentioned is for a specific regime of physical parameters corresponding to the experiment \cite{Park2020}. Here as we are focusing on protocol, though interesting, we will not be focusing on the exact behavior of the noise as a function of internal parameters.

\section{Methodology for distinguishing between states:}\label{sec:method}

It is worth noting that for a given interface in any state, although the actual values of CCC Fano Factors differ across different equilibration regimes, the overall behavior of the electrical shot noise (exponentially suppressed or constant) is preserved in all cases except in $\nu_i=3$ hosting the particle-hole conjugate states. Both Anti Read-Rezayi and BS-aPf states exhibit exponentially suppressed noise in full thermal equilibration but show a constant non-zero CCC Fano-Factor in a partial thermal equilibration regime. This flip is only seen when $\nu_i=3$, leading to one of the more important advantages of our methodology to distinguish the states because the $\nu_i=3$ interface acts as a clear indicator of the thermal equilibration regime. 

A significant portion of our protocol relies on looking at the presence and absence of noise, which is much easier than having to rely on the actual measurements of the noise every step of the way. Another way this protocol begins to prove useful is the fact that one does not need to rely on the knowledge of the true thermal equilibration length of the sample given the limited studies and numerical results in the case of the $\nu = 12/5$ FQH state. {One should be mindful of the fact that a key assumption on which this protocol is designed is that the device is always charge equilibrated. Previous experimental studies on $\nu=5/2$ \cite{Dutta2022, Srivastav2021, Melcer2022} show charge equilibration length to be 28 $\mu$m at 10 mK temperature in GaAs conductor and even lower in Graphene based quantum Hall systems. Complimenting these results, further works \cite{Dutta2022_Iso} show that the thermal equilibration lengths for $\nu=5/2$ lie at about 160 $\mu$m at 11 mK temperature in GaAs. Therefore we believe that the assumption that extending this pattern of charge always being equilibrated even for $\nu=12/5$ is quite reasonable and backed by experiments.}

With this prologue let us start the description of the protocol. Looking at \cref{Table} we can make the interesting deduction that irrespective of the thermal equilibration regime, if the noise detected in the $\{12/5,7/3\}$ interface is zero then the only possible candidate state the system can be in is Haldane-Heirarchy. This means that we can successfully determine if the system is in HH state just by knowing if the contacts are noisy or not (the first decision point in \cref{fig:flowchart}). After having checked for HH, if it turns out that the noise is not zero, we can safely eliminate the possibility of HH, which also limits the number of possibilities making it easier to look for patterns. Once again, if the noise in the $\{12/5,2\}$ interface is zero, irrespective of the thermal equilibration regime, we can conclude that the FQH state is the Bonderson-Slingerland state with $N=1$ (second decision point in \cref{fig:flowchart}). It is important to note that both particle-like states are universal in their behavior across equilibration regimes just based on the presence and absence of noise. Both of these states are also clearly distinguishable based on just one simple measurement for each state. However, the remaining Particle-hole conjugate states do not share these properties. They are particularly hard to distinguish because there seem to be no distinguishing based directly on the measurements of noise, as described above these states also do not behave similarly across equilibration regimes when $\nu_i=3$. So, to proceed, we find out the equilibration regime in play by checking noise at $\nu_i=3$. If the system is fully equilibrated, the only way to distinguish between the states is by looking at the ranges of measured noise. BS $(N=-3)$ $(\nu_i=2)$ in full equilibration shows almost twice as much noise than A-RR in the same configuration, making the comparison a fairly reasonable question. In case the noise in the $\nu_i=3$ interface is non-zero, there no longer exists any reasonably large difference in FFs in any configuration to distinguish it based on one measurement alone. Here the ratio of FFs at $\nu_i=7/3$ to $\nu_i=3$ is $\approx 2.9$ for A-RR state and $\approx 1.9$ for BS $(N=-3)$. Like in the previous case, it turns out that the difference is respectable enough to allow for comparison. We have chosen to make the comparison limit to be the median of the two values to accommodate for experimental differences in individual FF values. An important remark we make here is that the numerical calculations for the current-current correlations might sometimes be dependent on the value of $\gamma$ as shown in \cref{eq:noiseIntegral}. But this does not affect the above discussed protocol because only the Anti-RR and the BS-MR $(N=-3)$ states of the $\{\frac{12}{5},\frac{7}{3}\}$ and $\{\frac{12}{5},2\}$ interfaces are affected by changing $\gamma$. We have considered $\gamma$ to be unity to be consistent with \cite{SM5by2}, but this might not always be the case. One can see by careful analysis that this does not affect the distinguishability of the states as suggested by \cref{fig:flowchart}. In the worst-case scenario, if both fano-factors are close and comparing ratios is no longer feasible, one can look at noise scaling methods as discussed in \cite{Park2024}.

% We would like to end this discussion by recalling a key assumption 

\section{Outlook \& closing remarks}
\label{sec:outlook}
Given the immense interest in creating the topological qubits, the dark horses are the fractional quantum Hall states. In that journey, the 12/5 is a very strong candidate as this provides the possibility of universal quantum computing \cite{RevModPhys.80.1083,stern2010non,freedman2000modularfunctoruniversalquantum,PhysRevLett.95.140503,PhysRevB.75.165310} due to the possibility of Fibonacci anyons \cite{PhysRevB.59.8084}. However, we need a protocol to know which ground state it is without ambiguity. Recently \cite{PhysRevB.108.L241102} also attempted to classify candidate states of the $12/5$ FQH state using shot noise, by looking at the central charge. Their proposed protocol needs very fine control over length scales such that the device
is in full charge equilibration but remain unequilibrated in the thermal sector. However, equilibration length scales are not controllable parameters and come from the details of the device, and are also not well known in the case of $12/5$. Thus, we take a more cautious approach where we need a protocol that allows to distinguish among all candidates without knowing the thermal equilibration length. We also work for full and partial cases, as that will allow the devices to be larger, thus making the device easier to fabricate.
    There have been several experiments in recent times that measure the cross-correlations \cite{lee2023partitioning, ruelle2024timedomainbraidinganyons, PhysRevX.13.011031} and similarly control other very important parameters like heating of the system and contact resistance. We have seen that it is possible to maintain a base temperature of the system reliably throughout the measurement \cite{Melcer2022, Srivastav2021, roy2025halfintegerthermalconductanceabsence}. Thus our proposal is realizable in the current state of the art facilities.

Our protocol shows the power of noise-based investigations and that they can be extended from $5/2$ to more `complicated' filling fractions like $12/5$. We also show how much, without knowing the detailed number of the fano factor, we can still make the distinctions via qualitative differences. These models are also experimentally feasible \cite{Dutta2022_Iso, Dutta2022, Cohen2019, Biswas2022, Hashisaka2021, Hirayama2023} as long as we can establish the filter geometry via gating. Also, we must realize that in \textit{graphene} quantum Hall, the possibility of similar states is much higher due to the valley and the particle type hole type filling present in graphene \cite{Chandan2018SSP, PhysRevB.98.155421, Paul2022}. These phases might also be studied using the same principle described here.

\begin{acknowledgements}
We thank Sourav Manna for illuminating discussions. AD was supported by IISER, Tirupati Startup grant and ANRF/ECRG/2024/001172/PMS.
\end{acknowledgements}

\bibliography{refrence}

\end{document}